\documentclass[12pt,prd]{revtex4-1}
\usepackage{graphicx} 

\begin{document}
%\input epsf.tex    %<-If you need EPS figures to be
                   %  called in {figure} environment for PC
%\input epsf.def   %<-If you need EPS figures to be
                   %  called in {figure} environment for Macintosh

%\input psfig.sty

%\jname{}
%\jyear{2014}
%\jvol{}
%\ARinfo{}

\title{ TeV-Scale strings}
\markboth{David Berenstein }{TeV-Scale strings}

\author{David Berenstein}
\affiliation{Department of Physics, University of California, Santa Barbara, CA 93106}


\begin{abstract}
This article discusses the status of string physics where the string tension is around the TeV scale. The article covers model building basics for perturbative strings, based on D-brane configurations.
The effective low energy physics description of such string constructions is analyzed: how anomaly cancellation is implemented, how fast proton decay is avoided and how D-brane models  lead to additional $Z'$ particles. This review also discusses direct search bounds for strings at the TeV scale, as well as theoretical issues with model building related to flavor physics and axions.
 \end{abstract}

\maketitle

\section{INTRODUCTION}

The recent discovery of the Higgs particle with a mass of about 126 GeV at the LHC \cite{Aad:2012tfa, Chatrchyan:2012ufa} has made the Standard Model of Particle Physics complete, at least from the point of view of the electroweak interactions. We now have an experimentally verified weakly coupled and perturbatively renormalizable UV theory of the electroweak interactions as originally described by Weinberg \cite{SM}. Neutrino masses can be accommodated in this  scheme if there are right handed neutrino species, which can give a small mass to the neutrino via a see-saw mechanism, or they can have a finely tuned Dirac mass. In this sense there is no theoretical need for more particles, except for the fact that renormalization group running produces a Landau pole in the UV, far beyond any scale that can be tested in accelerator experiemnets.

The Standard Model of particle physics ignores the problem of gravity and it's possible quantum completeness.
A complete theory of nature should describe how to deal with gravity (in the strong gravity regime) and it should also explain the origin of dark matter. Dark matter has so far only been detected by its gravitational effects on matter. Previous claims of direct detection of light dark matter have not withstood the test of time: the LUX experiment has refuted those claims  soundly \cite{Akerib:2013tjd}.

This is where we stand today: we have the Standard Model, we have gravity, and we have very little guidance from experiment on what dark matter is made of. The clues we have from experiments and observation for physics beyond the Standard Model point to scales that are inaccessible, although the WIMP miracle argument suggests that the physics of dark matter is around the corner.  Neutrino masses suggest the existence of right handed neutrinos below about $10^{15}-10^{16}$ GeV, while the axion bounds suggest an axion decay constant, if an invisible axion exists, that is  larger than about $f_a > 10^{10}$ GeV.

To make progress we need to address the hierarchy problem: the dimensionless number relating the cosmological constant to the Planck scale involves a fine tuning of order $\Omega M_p^{-4} \simeq 10^{-120}$. This is the worst offender in terms of being an {\em unreasonable} small number: not of order one, but there are others. We call this hierarchy  a problem because is not stable under the renormalization group flow. We can not argue for its naturalness based on a symmetry principle \cite{'tHooft:1979bh}. The best {\em solution} to the cosmological constant problem, if it can be called such,   is based on anthropic arguments \cite{Anthropic} and the existence of a large number of vacua in string theory \cite{Douglas:2003um} (much larger than $10^{500}$). 

This is the place where strings at a TeV scale fits into the picture of naturalness: one tries to eliminate the appearance of additional high energy scales in nature and one asumes that the physics of quantum gravity appears close to the electroweak scale and in particular, that quantum gravity theory might be made of strings.
If string states can have masses of order the TeV scale, there would be very many new particles (stringy resonances) appearing at accelerator experiments \cite{Lykken:1996fj}. Analyzing such possibilities and their detailed phenomenology is interesting. Even black holes could be produced in accelerator experiments \cite{Giddings:2001bu,Dimopoulos:2001hw}.

String physics at the TeV scale is not a well defined subject. I will proceed  to narrow the definition in order to be able to fit a coherent description of what a status of strings at the TeV scale can be described as separate from other phenomenology models.

String theory is a  theory where the fundamental particles in nature are replaced by the quantum excitations of a fundamental string. The superstring theory in 10D is anomaly free \cite{GS} and is argued to be a perturbatively finite theory of quantum gravity. It is compatible with supersymmetry and Grand Unification. 

With the advent of string dualities \cite{Witten:1995ex},  the five perturbative descriptions of string theories in 10D are unified into a grander scheme where all of the string theories are limits of a single non-perturbative theory called M-theory. They are connected to each other by passing through a strong coupling.
With the gauge/gravity duality \cite{Maldacena:1997re} not only can the dimensionality of physics depend on the duality frame, but string theory compactifications can be completely equivalent to quantum field theories at strong coupling. The lines between what constitutes string theory and quantum field theory research have been blurred to the point that it is hard to say what constitutes research in string theory and what does not.

For the purposes of this article, we will narrowly define string physics to mean that the particles we see are actual string excitations (not D-branes nor any other non-perturbative effect) and the  string theory dynamics  is perturbative at the string scale: the coupling constant controlling the splitting and joining of strings is small.  We will furthermore require that the string scale is below about $10^5$ TeV.

 Perturbative heterotic  string phenomenology puts the string scale close to the GUT scale of about $10^{15}$ GeV, and the Kaluza-Klein excitations above $10^7$ GeV \cite{Caceres:1996is}.
 For the string scale to be low and the physics to be described by perturbation theory it is necessary to use type II string constructions, where the dynamics of gauge interactions and matter is described by open strings ending on Drichlet-branes  \cite{Dai:1989ua, Polchinski:1995mt} or D-branes for short. 
 
 This review will cover constructions with D-branes and low string scales. Also constructions that are inspired by the effective field theory of D-brane setups. We will also discuss the phenomenological implications for accelerator experiments and phenomenological bounds on such physics.
 
Given the constraints on the size of the paper and the number of references that can be included, it is impossible to cite all the works that have contributed to this body of knowledge, nor to be as detailed in the exposition as might have been desired. 
Many of the cited works refer to  specific examples, and they can serve as an illustration of the concepts, but not necessarily as the sole point of origin of an idea. The author apologizes for having to leave out many such works from the references that are cited.
To address these issues, many references are to other previous reviews. The reader should consult such works to get a more complete picture of all the literature on the subject.

\section{D-BRANES AND BASIC MODEL BUILDING}

\subsection{D-branes}

To describe D-branes, it is important first to describe how the type II strings operate. The reader should consult a standard textbook for details \cite{Polchinski:1998rq,Polchinski:1998rr}. The basic idea of string theory is that one can define an S-matrix order by order in perturbation theory by performing a path integral over all 2-dimensional Euclidean closed surfaces embedded in some space-time manifold ${\cal M}$. Naively, the action of such a world sheet is just the volume of the embedded surface and this would correspond to the Nambu-Goto string.\begin{equation}
S_{ws}\simeq \frac 1{2\pi \alpha'} \int d^2 \sigma \sqrt{\det(g_{\mu\nu}(X) \partial_\alpha X^\mu \partial_\beta X^\nu)}=  \frac 1{2\pi \alpha'} \int d^2 \sigma \sqrt{\det(g_{\alpha\beta})}
\end{equation}
where $g_{\alpha\beta}$ is the induced metric on the world sheet. And $(2\pi \alpha')^{-1}\simeq \ell_s^{-2}$ is the string tension, which is needed to restore units. 

To properly define the path integral one needs to include a world sheet dynamical metric $\gamma_{\alpha\beta}$ and Weyl invariance  \cite{Polyakov}. Quantum consistency  requires conformal symmetry at the quantum level: the theory is at a fixed point of the renormalization group. This governs the properties of ${\cal M}$: ${\cal M}$ satisfies the Einstein equations coupled to matter. Moreover, one has a critical condition where one can determine the space-time dimension of $\cal {M}$. This gives $D=26$ for flat space in the bosonic string. Given such a background, one can then in principle find the spectrum of strings propagating in such a geometry. The main result is that the string spectrum in flat space gives particles of various spins and masses, but it is mainly characterized by Regge behavior: the spectrum of masses versus angular momentum arranges itself in straight lines
\begin{equation}
M^2 \simeq 4 (\alpha')^{-1} J+a_0
\end{equation}
for various values of $a_0$. Only particles of spin less than or equal to two can be massless in this setup.

 For general ${\cal M}$ and at large volume,  one can also show that deformations of the metric of ${\cal M}$ and the matter content are realized by excitations of the string itself. To obtain space-time fermionic degrees of freedom one needs to introduce fermions on the string world sheet. The best way to do so in practical terms is via the Neveu-Schwartz-Ramond (NSR) formulation of the string, where the world sheet theory on the strings needs to have world sheet supersymmetry. In this case the critical dimension is $10$.  Supersymmetry on the worldsheet requires fermion partners to the $X^\mu$ bosons, which are usually called $\psi^\mu$. In the NSR formulation of the string, these fermions can have various boundary conditions on a circle. If they are anti-periodic, 
 we call it the Neveu Schwartz sector (NS), and if they are periodic we call it the Ramond sector (R). It is in the second case that the fermions can have zero-modes on a circle. The quantization of these zero modes gives rise to a Clifford algebra and the set of ground states are an irreducible representation of this Clifford algebra: these are spinors.  
  In the type II string, these world sheet fermions can be both left movers or right movers. The boundary conditions need to be set independently for left and right movers. There are now four sectors possible: NSNS, NSR, RNS,RR. The first and last give rise to space-time bosons (the Ramond-Ramond sector representation under the Lorentz group is the product of two spinors).
Consistency requires that the Ramond sector have only one possible chirality. If the chiralities of the NS-R and R-NS sector are opposite, 
then we call the theory type IIA, and it is a non-chiral ten dimensional theory of gravity with 32 space-time supersymmetries (this is $N=8$ SUSY in four dimensional field theory conventions). If the chiralities of the NS-R and R-NS sector are the same, the theory is called type IIB and it gives rise to a chiral ten dimensional theory of gravity with 32 supersymmetries. Massless states come from all four sectors. The RR-sector gives rise to  generalized versions of electromagnetism for higher dimensional objects.
The basics for computations of amplitudes requires a rather detailed understanding of conformal field theory on the world sheet \cite{Friedan:1985ge}, especially at higher loop orders (see \cite{Witten:2012bh} for recent progress on this issue) and
is beyond the scope of the present article. 

The idea of D-branes \cite{Dai:1989ua, Polchinski:1995mt} is that in these theories one can introduce topological defects where strings are allowed to end. Such an end to a string puts a boundary condition on the world sheet degrees of freedom: it can turn a left moving degree of freedom into a right moving degree of freedom. For bosons, having a fixed end is a Dirichlet boundary condition. Hence, the D-branes get their D prefix from this observation.
Such a boundary condition can preserve some of the supersymmetry. For the type IIA string, the even D-branes are supersymmetric (D0, D2, etc), while for the type IIB string it is the odd D-branes that are supersymmetric (D-1, D1, D3, \dots). The convention here is that a Dp-brane has p spatial dimensions and one time direction. A D-1 brane is an instanton: it has no extension in time.

The open strings ending on a Dp-brane can also be quantized and are soluble in flat space for a flat brane.  In particular, D-branes support massless spin one particles on their woldvolume: these are gauge fields. These also have Regge behavior, where 
\begin{equation}
M^2 \simeq (\alpha')^{-1} J + b
\end{equation}
and the slope of the Regge trajectories is different. The only massless particles are of spin less than or equal to one.

For weakly curved D-branes, one can show that the full effective action of the D-brane is a nonabelian generalization of the Dirac-Born-Infeld action \cite{Leigh}. 
Notice that open string excitations can only carry momentum in the directions parallel to the brane: it's endpoints are not allowed to move away from the brane: the effective field theory of strings at low energies that ends on the D-brane lives in lower dimensions.

One can have more than one Dp-brane, parallel to each other. 
In the limit of zero distance, $N$  D-branes can be stacked on top of one another and the number of massless spin one degrees of freedom is $N^2$. These give rise to a non-abelian Yang Mills action at low energies.  In the case of N branes on top of each other one gets a $U(N)$ gauge theory. This is the case for oriented strings: where there is a clear left and right direction on the string, so each endpoint of the strings knows this information. Strings can only split and join if they end on the same brane, and if the orientations of the string match. The joining of strings is like matrix multiplication and defines a generalized version of such matrix multiplication, called a star product, which can lead to a formulation of open string field theory \cite{Witten:1985cc}.
A left endpoint will have a fundamental charge of the $U(N)$ and a right endpoint will have an anti fundamental charge. This can be thought of as remembering on which brane each string ends and with which orientation. 
Strings stretching between different stacks of branes will be in bifundamental representations: fundamental under one stack and anti-fundamental with respect to the other and can have various spin quantum numbers.  A good general reference on D-branes is the textbook by C. V. Johnson \cite{Johnson:2003gi}

One can also get $SO(N)$ and $Sp(N)$ groups if one allows unoriented strings. This requires a further projection called the Orientifold projection: this is an identification between left and right movers on the world sheet (one gauges this discrete $Z_2$ symmetry). This is usually handled by a method of images.
 This projection deletes the information of orientation, so although each endpoint carries a fundamental charge, it can be either fundamental or anti fundamental: this depends on details of the projection.
These orientifold  identifications can have fixed points giving rise to quotient singularities in ${\cal M}$. If one places D-branes at a fixed sub-manifold under such a projection, one can obtain these other matrix multiplication groups as gauge groups. 
One can also do other types of identifications 
giving rise to quotient singularities called  orbifold spaces \cite{DHVW}. These keep the orientation of the string.  All of these  singularities give rise to consistent string theories with a well defined process for computing scattering amplitudes. They are not to be treated as singular: only the naive space-time metric and topology of ${\cal M}$ will look singular. Since a string segment always have two ends, all open strings are in generalized bifundamental representations: they are fundamental or antifundamental at each end of the string. This allows the introduction of symmetric and antisymmetric 2-tensor representations of $U(N), SO(N), SP(N)$. It is possible to get other groups at strong coupling, but those setups are beyond the scope of the present article.

 The coupling of open strings to each other is controlled by a coupling constant $g_o$. The coupling to closed strings is
determined by this open string coupling constant, $g_c\simeq g_o^2$. This coupling constant is determined by the expectation value of a string field, called the dilation, $g_c^2\simeq \exp(- 2\phi)$. All of these coupling constants are measured at the string scale $\ell_s$. The ten dimensional gravitational constant is then $G_{10} \simeq g_c^{2}\ell_s^8 \simeq \ell_p^8 $, and one can make $\ell_s>> \ell_p$ by taking $g_c\to 0$. This is, in the perturbative regime, the string length is much longer than the Planck length. One has introduced a hierarchy of scales this way. The hierarchy is determined dynamically: it depends on the expectation values of the dilaton.  

Upon compactification, the properties of the low energy theory will depend on the details of the compactification. The shape of ${\cal M}$ can depend on continuous parameters called moduli. Each of these acts as the analogue of a coupling constant in four dimensions. The same is true with the embeddings of D-branes on ${\cal M}$. These moduli need to be stabilized: there is no cosmological evidence of  variations of the coupling constants observed in nature with time. Potentials for moduli  can be generated by a variety of effects, perturbative and non-perturbative. The details are beyond the scope of the present review. The number of vacua computed as the number of critical points of the moduli potential can be very vast \cite{Douglas:2003um}. Each of these can be considered as a different theory with different coupling constants for four dimensional observables. This set of vacua is called the landscape. The set is so large that it is impractical to list them: there are more such vacua than the number of photons in the visible Universe.

There is a lot of variety in this set of vacua, so there is no unique prediction from string theory that covers all of the possibilities. 
The most universal prediction, that is valid for all string theory vacua in a perturbative regime,  is the existence of Regge behavior at high energies controlled by the string tension. The string tension itself is a free parameter as far as low energy physics is concerned and various bounds can be put on it that are model dependent (this is, they depend on the type of string background geometry and data that is chosen). The non-perturbative regions are less well understood, but cover most of the parameter space.

\subsection{Large extra dimensions}

Now that we have the basic ingredients of string theory and D-branes, we can ask how to determine the low energy four dimensional physics. Assume that we have a compactification of  type II strings of the form ${\cal M}= R^{3,1}\times X$ where this has the product metric, where $X$ is a six dimensional manifold. Assume furthermore that there are D-branes (possibly of various dimensions)  wrapping various cycles on $X$. Call these cycles $C_\alpha$, where $\alpha$ is a label, and it is assumed that the cycles $C_\alpha$ are topologically non-trivial: a brane wrapping that cycle can not be deformed continuously to zero size. Branes wrapping each $C_\alpha$ will give rise to different gauge groups, which we will also label by $\alpha$. Depending on if we are working in type IIA or type IIB string theory, and if we want supersymmetry the cycles are either  Special Lagrangian 3-cycles (for type IIA), or holomorphic cycles (for type IIB). A more detailed description can be found in \cite{Ooguri:1996ck}

 The effective action for massless states in 10 D and the corresponding massless states on the D-branes is given roughly by
\begin{equation}
\int_{\cal M} \exp(-2\phi) \sqrt g R + \sum _{\alpha} \int_{C_\alpha}  \exp(-\phi) \sqrt g_{ind} (F^\alpha_{\mu\nu})^2+\dots
\end{equation} 
all measured in string units.
The first term is the Einstein-Hilbert action. The second sum of terms are the low energy limit of the (generalized non abelian) DBI action for the D-branes, when we only consider the gauge field excitations. 
The term $\sqrt{g}_{ind}$ is the induced metric on the wolrdvolume of the Brane. 
Upon dimensional reduction to four dimensions, by taking constant excitations in the compactified dimensions  $g_{\mu\nu}, A_\mu^\alpha$ along ${\cal M}$ we find that 
\begin{eqnarray}
(G^4_N )^{-1} &\simeq&  (G_N^{10})^{-1} \hbox{Vol} ({\cal M}) \\
(g^\alpha_{YM})^{-2}&=& g^{-2}_0 \hbox{Vol} (C^\alpha)
\end{eqnarray}
The volumes of  ${\cal M}$ and $C^\alpha$ can be very different from each other. The simplest way to see this is to choose a single D3-brane moving on ${\cal M}$. This has support at a point, whose volume is by definition one. In that case $g^{D3}_{YM}\simeq g_o$. The volume of ${\cal M}$ can be very big. In that case the effective four dimensional Planck length becomes very short (the effective Planck scale energy becomes large), and makes low energy gravity decoupled.  On the other hand the field theory degrees of freedom on the D3 brane can still be interacting if we take $g_o$ to be small but finite. In the limit where the volumes of  some $C^\alpha$ are finite, but the volume of $\cal M$ is infinite, one can decouple gravity and engineer an interacting field theory in four (or other) dimensions.
This leads to a program called geometric engineering \cite{Katz:1996fh}, where one uses the geometric information in higher dimensions to study the field theory dynamics of the engineered theory, and one solves geometric problems in higher dimensions to solve various details of the low energy dynamics of the field theory.

To have the string scale near the TeV scale, we need a large volume for ${\cal M}$ in order to fit the four dimensional Planck scale correctly. This can be disastrous for low energy phenomenology. The Kaluza-Klein excitations of charged particles could be very light and would be inconsistent with data: they are subject to the LEP bounds. Because the excitations on the D-brane live in lower dimensions, their Kaluza-Klein spectrum might have nothing to do with the scale of ${\cal M}$, only with the scale of $C^\alpha$. Hence, since these are the only degrees of freedom that carry charges under the low energy gauge groups that make up the Standard Model, they would not be produced in accelerator experiments because their mass would be out of reach. 
This can evade bounds on light charged particles \cite{ArkaniHamed:1998rs, Antoniadis:1998ig}. The problem fields associated to the Kaluza-Klein spectrum on ${\cal M}$ would be made only of closed strings: these are neutral particles and couple to ordinary physics very weakly,  with gravitational couplings. This is basically due to dilution of the  wave function of the fields on ${\cal M}$, which requires a factor of $\hbox{Vol}(M)^{-1/2}$ to have normalized fields, not necessarily to the smallness of the closed string coupling constant $g_c$. 

This solves the hierarchy problem between the string scale and the electroweak scale: they are of the same order of magnitude. The same is true for the ten dimensional Planck scale. However, it replaces it by the problem of why the extra dimensions are so large. The hierarchy problem becomes a geometric problem, where other tools of quantum field theory and classical gravity might give some insight.

Notice that for models based on this type of analysis, all large extra dimension bounds from accelerator experiments (as in \cite{Mirabelli:1998rt}) are bounds on {\em stringy} phenomenology as well. The same is true for supersymmetry, as many of these constructions can be made supersymmetric. Since in principle neither large extra dimensions nor supersymmetry are predictions that are unique to string theory, one needs to go further to find evidence for string theory to be able to call it such. In general it turns out the string theory models are more restrictive: there are relations between couplings that are not {\em just effective field theory}. These give additional tests that can be used to make claims about stringy physics being due to strings. 

\subsection{Warping and throats}

The next variation on basic model building is the realization that product geometries as described in the previous section are not the only ones that are compatible with Lorentz invariance in four dimensions. One can also choose geometries where the topology is a product of $R^{3,1} \times {\cal M}$, but where the metric takes the form
\begin{equation}
ds^2 = \exp(-2f(y)) (dx_{3,1}^2) + ds^2({\cal M})
\end{equation}
where the $y$ are coordinates on $M$ and the $dx^2_{3,1}$ is the usual Lorentz metric in four dimensions. The function $\exp(-f(y))$ is called the warp factor. This means distances in four dimensions depend on the location of an object in the higher dimensions. These distances are measured between fixed values of the coordinates in $R^{3,1}$ and orthogonally to ${\cal M}$. As such, the notion of energy scales depends on the location of an object in higher dimensions. This is the familiar redshift in gravity and does not indicate a disparity between the local string scale and the local 10 dimensional Planck length. It indicates that there is a variation between the effective four dimensional Planck scale  (which is a global definition) and the local string scale, which needs to be corrected for redshift. This is a factor of $\exp(-f(y))$. This is familiar in the gauge/gravity correspondence \cite{Maldacena:1997re}.

These geometries are necessarily not Ricci flat, and need matter to support this curvature. If there are D-branes, they can themselves act as sources of the matter fields that bend the geometry. These matter fields deform the geometry in which the D-brane is embedded. Doing this procedure systematically is called including the back reaction of branes. A single D-brane has a tension of order $1/g_s$ \cite{Polchinski:1995mt}, not $1/g_s^2$ as would be naively expected for solitons. This is actually expected from the asymptotic behavior of the string perturbation series \cite{Shenker:1990uf}. For practical purposes this means that back-reaction can be ignored sometimes, band it can be tuned  by changing the number of branes in a stack. 

For a stack of Dp-branes, the back-reacted supergravity solutions are well known \cite{Horowitz:1991cd}: they are extremal black p-brane solutions. They can have a near horizon region where there is an infinite redshift. The local string scale can be arbitrarily small depending on the location of an object in the ten dimensional geometry. One can take a low energy limit by focusing on this high redshift region. This observation leads to the gauge/gravity duality \cite{Maldacena:1997re}. Which description is better depends on the 't Hooft coupling constant  \cite{'tHooft:1973jz} of the gauge theory: the gravity description is semiclassical  at strong 't Hooft coupling.

This provides a different avenue to solve the hierarchy problem: one uses redshifts to lower the electroweak scale relative to the four dimensional Planck scale \cite{Randall:1999ee} and one can even allow for infinitely large extra dimensions \cite{Randall:1999vf}.
These redshifts might be generated by a stack of D-branes which is independent of the Standard model, or might give a gravity dual solution to technicolor theories (see \cite{Farhi:1980xs,Lane:2002wv} for  reviews of technicolor). In this case the techni-meson and techni-hadron resonances would give an effective string theory of the confined technicolor theory. Such a string theory is iessentially indistinguishable from the ten-dimensional strings of type IIA and type IIB strings.

Field theory in higher dimensions as required by these setups will need a UV completion: the theories are not perturbatively renormalizable. As such, one will eventually have to reintroduce strings or another device to make the models UV complete.

\subsection{Tadpole cancellation and anomalies}

An interesting question to ask at this moment is how we actually get chiral matter from stacks of branes. 
This is provided by how they intersect each other (see \cite{Berkooz:1996km} and references therein).  If  two stacks of branes intersect transversely (we can not remove their intersection apart from each other by small deformations), one can show that chiral massless fermions live in their intersection. Depending on the angles between the branes, they can also preserve a fraction of the supersymmetry.
Since chiral fermions are required to build the Standard model, this result shows that acceptable phenomenology from the point of view of
basic particle physics is in principle possible. Also, since the Standard model gauge group $SU(3) \times SU(2)\times U(1)$ can be embedded in product groups of the type $\prod U(N) \times Sp(M)\times SO(K)$, it is at least in principle possible to find a realization of the Standard Model within this class of models.

\begin{figure}
\begin{center}
\includegraphics[width=10 cm]{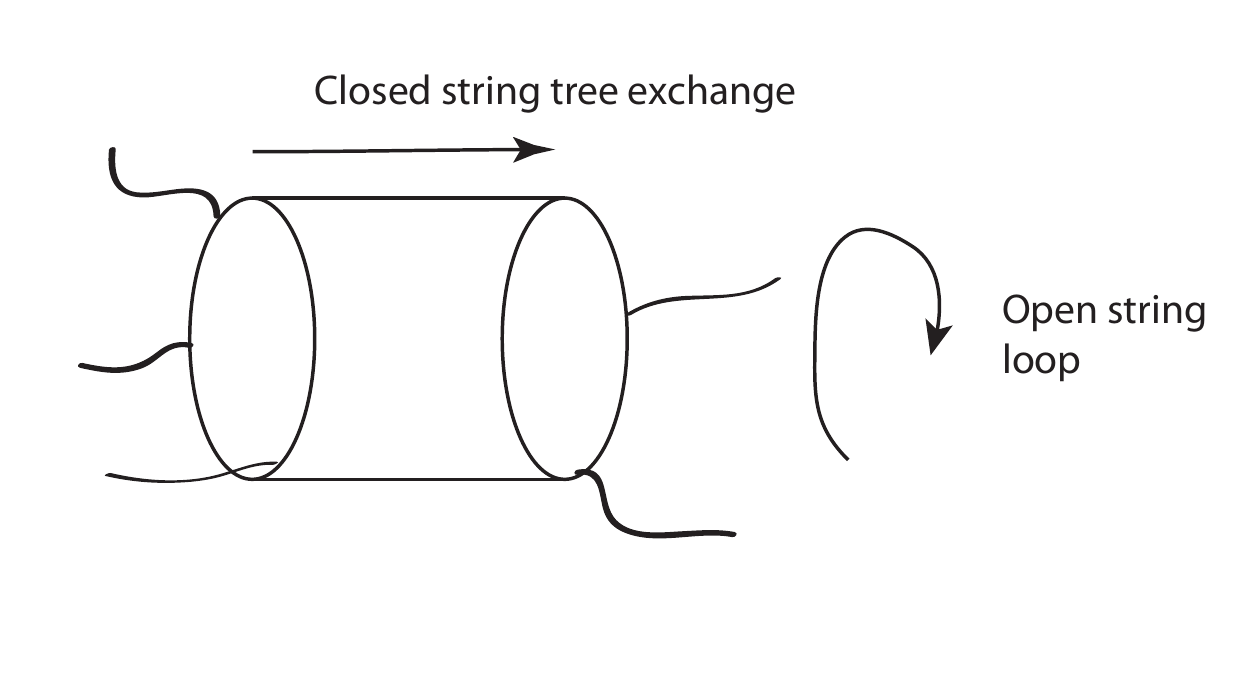}
\end{center}
\caption{A one loop of open strings can be viewed as a tree exchange of closed strings by cutting in the $s$ versus $t$ channels. The wavy lines indicate possible external particles attached to the loop.}
\label{fig:cylinder}
\end{figure}

Chiral fermions can produce anomalies via loop diagrams. Quantum consistency of a configuration requires that field theory anomalies are cancelled. A one loop diagram of open strings can have a field theory limit as a Feynman diagram with the fermions propagating in the loop, this configuration is a cylinder, and in the case of unoriented strings one can also have a Mobius strip. Finiteness of the cylinder diagram (or combination of diagrams) provides for anomaly cancellation. 
The anomaly produced by the matter fermion loop arises from a  region of integration over worldsheet parameters where the cylinder is a very thin strip tied into a loop. This is 
 balanced by a tree level contribution from closed string exchange: this is the region of the worldsheet parameters where the cylinder is very long .In the string diagram, both processes are part of the same  diagram: a cylinder. This is depicted in figure \ref{fig:cylinder}.
 
This is the so called Green-Schwarz mechanism. If the anomaly polynomial does not factorize, one of the  gauge groups will be anomalous, whereas if it factorizes, it is cancelled by a closed string exchange. From the point of view of the low dimensional effective action, 
the closed string state that participates in the anomaly needs to have tree level couplings to the gauge groups. 

One can analyze when the field theory anomalies cancel in various interesting setups \cite{Leigh:1998hj,Ibanez:1998qp} and find that the condition of anomaly cancellation can be derived from another consistency condition of string compactifications: tadpole vanishing. The idea of tadpole diagrams is very simple: D-branes are sources of RR fields, and indeed, they act as the fundamental charged objects under the generalized electrodynamics that they generate. When we put a gauge theory on a finite volume box, gauge invariance requires that the net charge vanishes. The same condition needs to be applied to setups where D-branes wrap various cycles: they are a source for the field strength in the compact directions and the charge needs to go somewhere. The tadpole constraints are in general stronger than anomaly cancellation and they imply the cubic anomaly cancellation \cite{Aldazabal:1999nu}. 

Factorization of anomalies means that the $U(1)$ sector of $U(N)$ can have mixed anomalies with some other gauge groups in four dimensions. The cancellation of these mixed anomalies is handled by the Green-Schwarz mechanism, and makes the $U(1)$ massive. The closed string state that also participates in the anomaly is a scalar and  has axion like couplings to the other gauge groups, while it has a shift symmetry under the $U(1)$ that has mixed anomalies. Gauge invariance of the kinetic term for this axion-like particle implies a Stueckleberg mass for the gauge field. This mass comes from open-closed string mixing, and is of order $g_o$ in string units. Such $U(1)$ gauge fields are massive, but might be much lighter than other typical strings.

\section{ TOP DOWN AND BOTTOM-UP APPROACHES TO MODEL BUILDING}

A top down approach to phenomenology proceeds from trying to find a globally consistent solution to a string theory compactification. One chooses a geometry $\cal M$ (usually a Calabi-Yau manifold in order to have some supersymmetry \cite{CHSW}) and a UV string theory (for most of this section we will take a type IIA string theory). Then we need to put supersymmetric branes that wrap various cycles in the 
geometry and verify the consistency conditions (anomalies, tadpoles, etc). To have branes and satisfy tadpole conditions plus supersymmetry this necessitates an orientifold projection. After these choices are made, one computes the matter spectrum, the coupling constants  (as functions of the moduli, and not always doable in practice). One then needs to solve the dynamics. For example, how SUSY is broken and how the Standard Model Lagrangian actually emerges. In practice this can not be done in full generality: we do not even know the metric of smooth non-trivial Calabi-Yau geometries.

\subsection{Grand Unification}

The simplest way to get the Standard Model field content from a single stack of branes is to consider a Grand Unified theory. For example, one could try to produce  a standard $SU(5)$ GUT, with chiral matter in the ${\bf 10}$ and ${\bf \bar 5}$ representations \cite{Georgi:1974sy}. These representations are not bifundamental. The bifundamental of $SU(5)$, $(\bf 5, \bar 5)\simeq \bf 24\oplus 1$,  is the adjoint representation, which is appropriate for the gauge group and some of the Higgs fields. 
Also, $SU(5)$ is not $U(5)$, but the $U(1)$ has mixed anomalies and can be made massive by the Green-Schwarz terms in the action.
The $\bf 10$ is an antisymmetric representation: this is in the product of two fundamentals, so such a construction necessarily requires an orientifold. For the ${\bf \bar 5}$ one needs another brane to end on. This can give rise to a $SO(1)$ or $U(1)$ gauge group.
A Higgs scalar can also be accommodated in the adjoint and the $\bf 5$. To have 3 families, the $U(5)$ stack needs to self-intersect with it's orientifold image three times.

It is possible to estimate the proton decay by certain high dimension operators which are independent of the model (see \cite{Klebanov:2003my} and references therein). A review of these results can be found in \cite{Nath:2006ut}. If one assumes a low string scale physics, and an even lower GUT scale,  such models are ruled out by current bounds on proton decay. 

One could also try to get an $SO(10)$ GUT model, but this is automatically ruled out: a spinor of $SO(10)$ is not in a binfundamental rtepresentation. Matter content restrictions of perturbative D-brane setups can remove certain  interesting phenomenological possibilities.

\subsection{Branes at angles in type IIA}

The next idea is to look for non-unified models. In such models one starts with many stacks of branes in some geometry and one tries to achieve three goals: get a gauge group that contains the Standard model, get matter content such that when appropriately broken at low energies to the Standard Model, the chiral states match those of the Standard model, and one needs to satisfy the tadpole constraints. Usually this is done with supersymmetry. After these goals are met, one needs to compute the couplings of matter and check if all the Standard Model couplings are produced by tree level interactions or not. One also needs to analyze the associated moduli fields and find how they are stabilized. The literature on the subject is very vast and this review can not make justice to all of these developments.
Instead, we refer the reader to the following reviews \cite{Blumenhagen:2005mu,Blumenhagen:2006ci,Marchesano:2007de}. These contain most of the  results up to 2008. The reader should note that the constructions of branes at angles in orientifolds of tori are essentially soluble models: in principle all the coupling constants can be computed (see \cite{Cremades:2003qj} for an  example of Yukawa couplings). This is also true for other ways of building exact world sheet Conformal Field Theories. 

For supersymmetric models (these are the ones that have been studied most), the branes at angles are all given by D6 branes wrapping cycles on a quotient of a torus, where all of the cycles need to be comparable in size, lets call it $L$. The gauge coupling constants in four dimensions are given by volumes of three cycles, thus they are of size $1/g_{YM}^2\simeq L^3/\ell_s^3g^2_o$, whereas the Planck length in four dimensions is given by $1/\ell_{4,p}^2\simeq L^6/ (g_c ^2 \ell_s^8)\simeq 1/(\ell^2_s g_{YM}^4) $, this is, the String scale and the four dimensional Planck scale can not be that different from each other, see for example \cite{Cvetic:2001nr}. To change that, and keep supersymmetry, we need to produce 3-cycles that are parametrically small with respect to the volume. This requires studying more general compactifications with  branes on Calabi-Yau orientifolds where the three volumes of different cycles are more easily adjustable. There are usually large numbers of closed string moduli on these geometries. Thus the problem  of moduli stabilization is worse. Moreover, constructing special Lagrangian 3-cycles is very difficult.

Introducing fluxes (which are allowed fields in supegravity) can stabilize most or all of the moduli \cite{DeWolfe:2005uu}. This generically happens in models where doing precise computations of the relevant phenomenology is very hard, if not downright impossible. Up to date, I believe that there are no known compact Calabi-Yau metrics for geometries that are not orbifolds.  
Developments related to analysis of such setups are reviewed in \cite{Grana:2005jc, Douglas:2006es} and involve a 
a lot of supergravity computations. 

Because for most of these models lowering the string scale to the TeV scale is  difficult, we will move to other setups where this is more easily done.

\subsection{Bottom up approach: Branes at singularities in type IIB string theory}

The idea of a bottom up approach is to take  seriously the statement that the large extra dimensions (the volume of $\cal M$) are large and that at the same time, the cycles on which the D-branes that the Standard Model is supported on are very small, essentially zero. We imagine that we can engineer a region where some non-trivial cycles have been collapsed to at or below the string scale essentially to a point. Such a region  is not locally a manifold: there is non-trivial  topology at a point. Therefore the `manifold' is singular in the location of the Standard model branes. This is a problem if we only have a na\"\i ve theory of gravity, but in string theory it is possible to have singular spaces from the classical point of view that are perfectly non-singular as far as the string interactions are concerned. The prototypical example is an orbifold \cite{DHVW} where one can actually compute the whole string partition function. In principle, this singularity can be attached to a variety of different compactifications, so the Standard Model physics does not uniquely determine one global geometry. Instead, one just looks at the physics near the singularity and one imagines that it is possible to paste this information on various different possibilities, which we don't assume we know. The low energy field theory predictions will not depend very much on this embedding. One can argue that closed strings decouple in this limit, as their wavefunction will be spread
over ${\cal M}$, so they effectively only couple gravitationally. The relevant dynamics  is the effective theory of open string states attached to the branes at the singularity. There are some exceptions to this rule: if a closed string mode is localized near the singularity, one can not throw it away.

 To simplify matters, one postulates that  the Standard Model is realized on a single point-like brane that is near the singularity, and that the moduli problem, as far  as this brane is concerned, has been solved. The local conditions near the singularity and the existence of the singularity determine the effective couplings and full dynamics in the low energy field theory on the brane. How the singularity came to be in the first place is not resolved but assumed.
To have a point-like brane in the transverse directions to the singularity and to have four dimensional field theory and supersymmetry, we need to require that a D3-brane is supersymmetric: this puts us into the type IIB string setup. Because of dualities, type IIB and type IIA are related to each other by mirror symmetry, so it is always possible --at least in principle-- to find a dual type IIA realization of these setups. However, if the type IIB string theory is at large volume, the type IIA string theory is not (this is most easily seen from the relation between mirror symmetry and T-duality \cite{Strominger:1996it}).

A calculable example of what the physics of open strings at a singularity can look like is provided by studying D-branes at orbfiolds. An orbifold for us will be a quotient of flat space by a discrete group $R^d/ \Gamma$, where $\Gamma$ is a discrete subset of the isometry group of $R^d$ that fixes the origin, $SO(d)$. 
 The technique to study D-branes on such a space was developed by Douglas and Moore \cite{Douglas:1996sw}. This is done by the method of images in the cover and such models are completely calculable.

The simplest way to describe the matter content is in terms of a decorated quiver or moose diagram. The nodes represent gauge groups and the arrows represent matter, this is depicted in figure \ref{fig:subquiver}. For orbifolds that preserve orientations of the string they are of type $U(N_i)$. The factors of $i$ are in one to one correspondence with irreducible representations of $\Gamma$. 
For a single brane in the bulk $N_i = \hbox{Dim}(R_i)$, the dimension of the representation of $\Gamma$ associated to the label $i$.

\begin{figure}[ht]
\begin{center}
$$ {\bullet \atop {U(N_1)}}\quad {\longrightarrow \atop{\psi_{12}}}\quad {\bullet \atop {U(N_2)}}$$
\end{center}
\caption{A subquiver diagram with two nodes and one arrow. The nodes are gauge groups and the arrow is in a bifundamental, with some spin labels and mass that are not indicated. The group theory repesentation is encoded by the orientation of the arrow. For unoriented setups one has to specify the orientations of both ends: this distinguishes the fundamental from the antifundamental representations of $U(N)$. }
\label{fig:subquiver}
\end{figure}

One can also change the values of $N_i$ to be more generic. A configuration with $N_i=1$ for some $i$ and $N_j=0$ for $i\neq j$ is called a fractional brane and  is stuck at the singularity. 

 If we further require that $\Gamma$ preserve some supersymmetry we are forced to consider  $\Gamma\subset SU(3)$. These are completely classified, and a table of the corresponding groups and quivers can be found in \cite{Hanany:1998sd}. One can not get the matter content of the Standard Model on the nose with a single brane in the bulk in an orbifold.  The best one can hope for with orbifolds is
 that we obtain a group with a quiver such that a sub quiver gives the Standard Model. One can look for models with 3 generations, and at least one three dimensional representation: this gives rise to a $U(3)$ color group. To get representations that have dimension greater than one, one needs to use a non-abelian orbifold. 
  
One then needs to follow a path to do phenomenology with these basic ingredients \cite{Aldazabal:2000sa}. A simple example is the quotient $R^6/\Delta_{27}$.  This is supersymmetric, with gauge group $U(3)_1\times U(3)_2\times U(1)^9$. One imagines that the weak theory  $SU(2)_W\subset U(3)_2$ and that hypercharge is a linear combination of the other $U(1)$ \cite{Berenstein:2001nk}. The matter content has three generations of left handed quark doublets: this is tied to the number of transverse complex dimensions.
One also finds that superpotential masses for leptons are forbidden, so in perturbation theory they need to be generated by higher dimension operators in the field theory, suppressed by the high energy scale (the string scale). To even think of making such a model viable, the string scale is necessarily low. One can show that such a model has a large number of chiral doublets, so it is not realistic. 

A quiver diagram as above does not determine a unique singularity.  One can consider orbifolds with discrete torsion \cite{Vafa:1986wx}. This introduces an extra twisting in how the orbifold is done, and it is realized by D-branes with projective representations of the group  $\Gamma$ \cite{Douglas:1998xa}. If one constructs examples of D-branes at singularities \cite{Douglas:1999hq,Berenstein:2000hy} one finds that various such orbifolds can give rise to the same quiver diagram, but the low effective field theory has different couplings. 

One can construct field theory quivers  associated to other singularities. The prototypical example involves strongly coupled superconformal  field theories in the infrared \cite{Klebanov:1998hh}, like the ones described by Leigh and Strassler \cite{Leigh:1995ep}.
It is interesting to note that if one adds fractional branes to this geometry, the effective low energy theory in the infrared is supersymmetric and confining. The effect of confinement, a non-perturbative effect, can be realized by changing the topology of the singularity and deforming the geometry \cite{Klebanov:2000hb} (see also \cite{Gopakumar:1998ki} ).

An interesting singularity to  study is the $dP_8$ singularity \cite{Verlinde:2005jr} which is closely related to the $\Delta_{27}$ quiver theory. This general setup is more interesting because it has more geometric moduli than orbifolds. 

In general one can deform the field theories obtained in various such ways to include other effective terms in the low energy action by adding fluxes, see  for example 
\cite{Camara:2003ku} to understand how some soft SUSY breaking terms can be generated. All such modifications can lead to arbitrary coupling constants on a quiver, so we can tune couplings to get the Standard model. It turns out that this is not quite right. String theory models are not just effective field theory: there are constraints. These go beyond gauge invariance. Some couplings are forbidden and some other couplings can only arise from higher orders in perturbation theory, therefore they are suppressed by powers of $g_c$.
If such constraints can be verified in nature one can claim to have evidence for strings.

\section{EFFECTIVE FIELD THEORY CONSTRAINTS}

\subsection{Effective Lagrangians generated by perturbation theory}

The author believes that the following are well known facts about the effective field theory constraints that arise from string perturbation theory, but it seems that they are not written down systematically in the literature. They can be inferred from standard textbooks, e.g. \cite{Polchinski:1998rq}.

A perturbative interaction of joining open oriented strings takes a pair  of quanta in a bifundamental representation $\Psi_a^b$ and $\tilde \psi_c^d$ and can join them into a new quantum $U_a^d$ or $\tilde U_c^b$ exactly if $b=c$ or if $d=a$. The ends of the two strings that  are joining are  in the same brane and the new string remembers the other ends of the strings that are fusing. An upper index indicates a fundamental representation, a lower index  an anti-fundamental representation.
 The labels $a,b,c,d$ can be charged under different gauge groups, and if a state is in a superposition of various of these, we have to sum over all components. 
  The interaction in field theory that would give rise to such a process would be of the form
 \begin{equation}
\sum \int A(a,b,d) \phi_a^b \tilde \phi_b^d \bar \Omega^a_d +B(a,c,d)  \tilde \phi_c^d  \phi_d^b \bar {\tilde \Omega}_b^c
 \end{equation}
where $\phi, \tilde \phi$ are the quantum fields associated to $\Psi,\tilde \Psi$, and $\Omega, \tilde \Omega$ are the fields associated to $U, \tilde U$ respectively. The numbers $A,B$ are the corresponding amplitudes for fusion. Notice that we need to conjugate $\Omega,\bar \Omega$ to distinguish incoming from outgoing particles in a Feynman graph. Each of $A,B$ is of order $g_o$.

We sum over all possible labels with amplitudes governed by gauge invariance. Also, a propagator would remember it's ends, so that 
$
(\Delta^{-1} )^a_b\! ^{b'}_{a'} \simeq \delta^a_{a'} \delta_b^{b'} (p^2+m^2)^{-1}
$
For unoriented strings the rules require a bit more care, but are essentially the same as above, supplemented by the orientifold projection. This projection  will identify the various types of ends and only allow particular superpositions of the labels in the cover.

When we integrate out a heavy open string field at tree level, we can generate new terms in the effective field theory.These are going to be of the form
\begin{equation}
\int  A^{i_1,i_2,\dots,i_n} (a,b,c, d, \dots,s)  (\phi^{[i_1]})_a^b ( \phi^{[i_2]})_b^d \dots (\phi^{[i_n]})^{a}_s
\end{equation}
where we always concatenate the vertices where the off-shell heavy string is connected  according to planar diagrams. We get diagrams that can only be described by a canonical order of fields, and canonical contraction of the labels, and $A^{[n]}\simeq g_o^{n-2}$: each external leg adds a power of $g_o$. This counting of powers of $g_o$ is the same as ordinary perturbation theory counting for tree level diagrams in $\phi^3$ field theory.

\begin{figure}[ht]
\begin{center}
\includegraphics[width=7 cm]{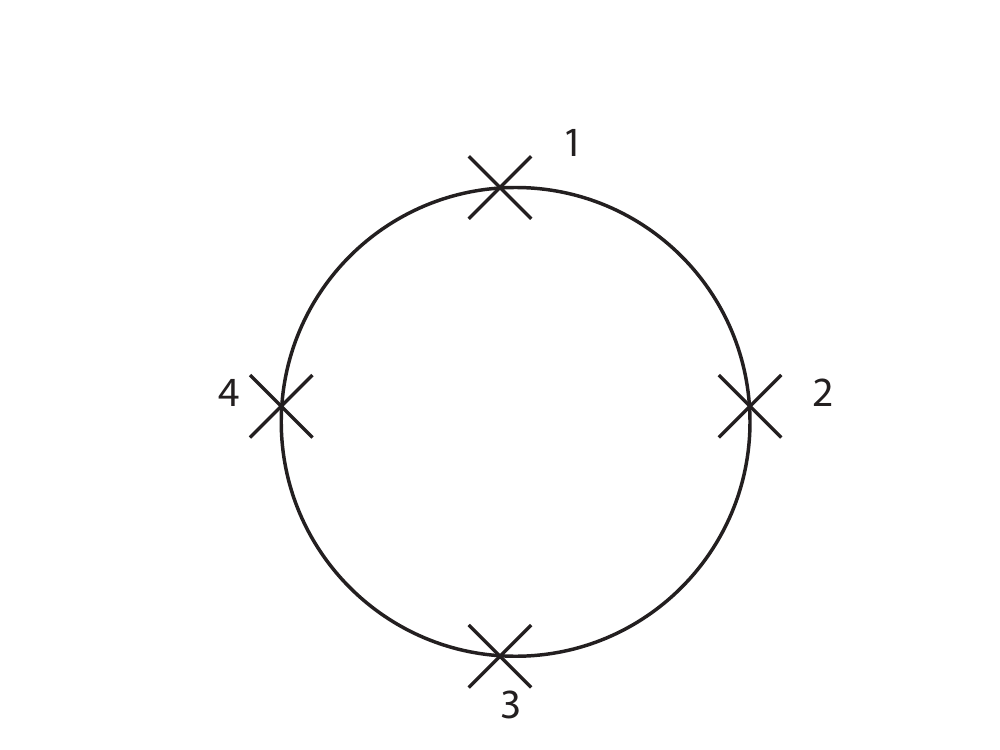}
\end{center}
\caption{A tree level open string diagram with four external fields $1,2,3,4$. We associate a Feynman rule $A^{1,2,3,4} Tr(\phi^{[1]}\phi^{[2]}\phi^{[3]}\phi^{[4]})$ to such a graph. Each edge of the disk needs to be ending on a particular D-brane stack. The open string insertions need to reflect how they end on the branes associated to each edge.
}
\label{fig:disk}
\end{figure}

All such couplings are a single trace effective action term in a quiver diagram. Single traces are a closed path in the quiver where the arrow ends are compatible.Tree level couplings of strings to each other are not completely arbitrary: they are of single trace type. These tree level diagrams are all obtained from a single disk diagram  with various external particle insertions at the edge of the disk. An example is shown in figure \ref{fig:disk}. The order of the particles around the disk indicates the order in which the trace is to be taken.

\begin{figure}[ht]
\begin{center}
\includegraphics[width=8cm]{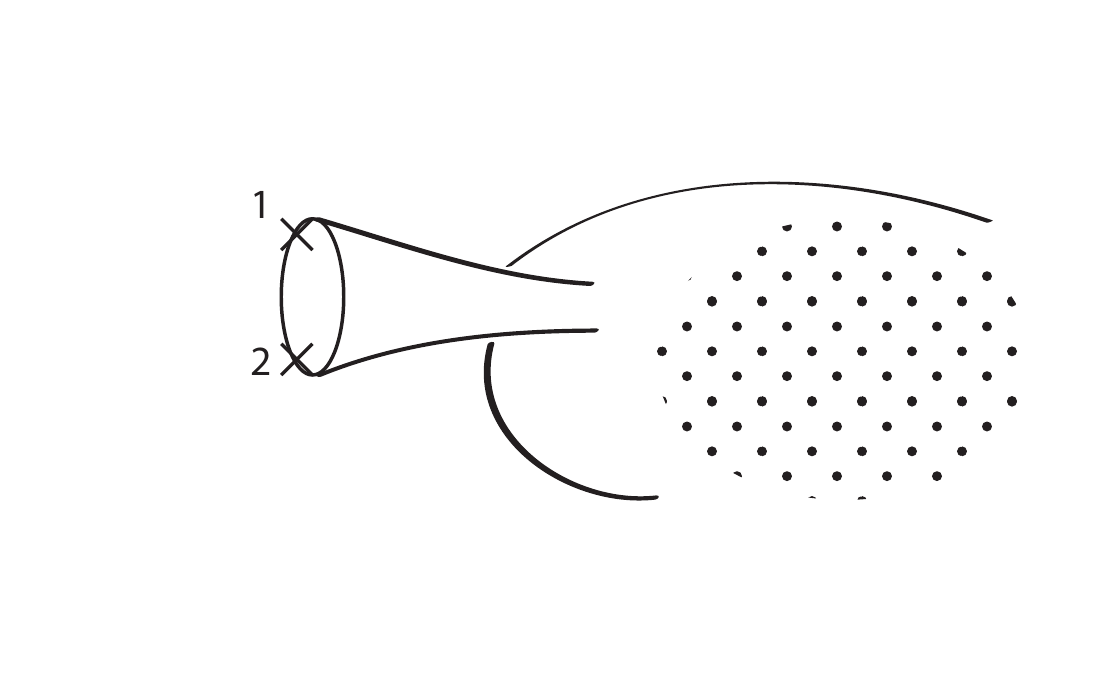}
\end{center}
\caption{Attaching a trace to another diagram: we add a power of $g_c^2\simeq g_c g_o^2$, one for the interaction of the closed string to the trace (a disk), and one for the interaction with the rest of the diagram. }
\label{fig:addtrace}
\end{figure}

When we go to loops of open strings, we can think of the corresponding diagrams as closed world sheets with holes. The power of $g_c$ that one associates to such a surface is determined by the genus. For each such hole in the world sheet we add a trace. For each trace we add, we need to add a an extra power of $g_o^2\simeq g_c$, which is easy to count if we imagine it as  higher tree level diagram with closed string intermediate states. This is depicted in figure \ref{fig:addtrace}. 
Multi-traces are therefore suppressed by additional powers of $g_c$ relative to single traces.  These coupling constants for multi-traces are to be thought of as being generated by loops. 

In a $U(N)$ gauge theory the $U(1)$ coupling constant and the $SU(N)$ coupling constant are the same at the level of disks. A difference between their coupling constants would be handled by a double trace, $[\hbox{tr}(F_{\mu\nu})]^2\simeq F_{U(1)}^2$. This would also occur for mixing kinetic terms between different $U(1)$ factors of stacks of branes.
Kinetic terms for the $U(1)$ gauge bosons are diagonal in a decomposition of the form $\prod U(N_i)$, and any mixing between the $U(1)_i$ kinetic terms is suppressed by loops.
Multi-traces are the only types of effective terms that are generated in perturbation theory.

\subsection{Forbidden perturbative couplings: the proton decay problem and generation of neutrino masses.}

The group invariants for $U(N), SO(N), SP(N)$ that are used to construct multi-traces are  built only from the $\delta ^a_b$ invariant of $SU(N)$. There is another invariant tensor in $SU(N)$: this is the $\epsilon_{i_1, \dots i_N}$ tensor. Coupling constants that use the $\epsilon$ tensor can not be generated in perturbation theory, because it is not a product of $\delta^a_b$ symbols.  Any such coupling constant in effective field theory is zero to all orders in string perturbation theory.  The easiest way to describe this is that in the $U(N)\to SU(N)$ breaking, by anomalies or just by making the corresponding $U(1)$ massive, the global $U(1) $ symmetry is a global symmetry of the perturbative effective action of the remaining fields.  

To generate such couplings in the dynamics one needs non-perturbative effects. These can be provided by Euclidean D-branes $E_a$. This gives a suppression to those couplings of order $\lambda_{np}\simeq \exp (- \hbox{Vol}( E_a))/g_c)<<1$ with possible additional suppressions and dependence  on moduli.

This reasoning makes all $\bf 10 - 10- 5$ couplings  in $SU(5)$  small and it also forbids the right handed quarks from being in antisymmetric representations of $U(3)$ \cite{Berenstein:2006aj}: the corresponding Yukawa couplings to the Higgs can only be generated non-perturbatively. If one needs to invoke some large non-perturbative effect to generate an important coupling for the Standard Model, like the top-Yukawa coupling, one can argue that this induces other large non-perturbative effects that make the perturbative analysis suspect. This usually leads to a ``disastrous proton decay" \cite{Kiritsis:2009sf}, although this can be ameliorated in flipped $SU(5)$ setups \cite{Anastasopoulos:2010hu}. 

In that case one is better off working in F-theory or M-theory where the $\bf 10 - 10- 5$ couplings can appear from geometric considerations. These are non-perturbative constructions and therefore are not suitable for what we have narrowly defined as strings at the TeV scale. How phenomenologically useful output arises in these setups can be found for example in \cite{Heckman:2010xz}. A recent review can be found in \cite{Maharana:2012tu}.  

Because of this, perturbative D-brane setups should be non-unified. This does not preclude a UV determination of the Weinberg angle. An example of such a determination is provided by requiring a non-trivial UV conformal field theory \cite{Frampton:2001xh}, or some other discrete symmetry. 

If one insists on the string scale being low, one would imagine that the proton decay could only be suppressed by the string scale, and this would be in conflict with the current bounds from experiment, where the lifetime of the proton is bound by $\tau>10^{31}$--$10^{33}$ years, depending on the mode of decay \cite{Beringer:1900zz}. This is not automatically the case. Since the $SU(3)$ gauge group is not of the form $U(N)$, it needs to be enhanced to a $U(3)$. This gauges Baryon number, a symmetry that is anomalous as above. The proton decay operators would use the $\epsilon$ tensor and thus are forbidden by the remnant global $U(1)$ symmetry in field theory \cite{Ibanez:1999it}. This result by Iba\~ nez and Quevedo is a consistency check that a low string scale is not automatically excluded by the long lifetime of the proton, and we will call the local to global symmetry conversion the Iba\~nez-Quevedo mechanism.    Proton decay operators would only be generated by non-perturbative effects.

In various models Lepton number is also gauged, although this is not mandatory. In such models one has to generate the neutrino masses non-perturbatively \cite{Blumenhagen:2006xt,Ibanez:2006da}. Computations of these instanton effects can be performed in some cases (such a systematic computation was initiated in \cite{Cvetic:2007ku}), and a general review can be found in \cite{Cvetic:2011vz}.  This can also affect the $\mu$ term in supersymmetric constructions. It is interesting to notice that not all of these instantons can be interpreted as field theory instantons \cite{Florea:2006si} and as such must be thought of as stringy in origin.

\section{GENERIC PREDICTIONS, TOY MODELS AND PHENOMENOLOGICAL BOUNDS}

We're finally ready to tackle the problem of phenomenological output for particle physics models. Some output is model dependent, some output can be considered generic and essentially model independent, but it still requires modeling in order to confront experiment. The first important statement is that the string scale is usually not predicted at all, but only indirectly. 

The second statement that is important to make is that up to date there is no construction of a realistic Standard Model or it's supersymmetric cousins where the problems of moduli-stabilization, computation of the low energy coupling constants and electroweak data including Yukawa couplings can be performed in detail. The biggest obstacle is that the details of Supersymmetry breaking have not been successfully carried out. One can still make orders of magnitude estimates  and argue for or against models based on these.

\subsection{Quiver arguments: $Z'$ particles, mutiple Higgs fields and  axions}

In general, if we want a brane stack with $SU(3)$ gauge theory for QCD, this gets enhanced to $U(3)$ at the string scale. This $U(1)$ can not be hypercharge: the leptons would not be accommodated in such a case, as they would also be colored. This is usually gauged baryon number. This first prediction is that there is at least one additional neutral $Z'$ particle \cite{Ghilencea:2002da}.  One then needs to decide if one is going to treat the $SU(2)$ of the weak interactions as a $U(2)$ or as a $Sp(1)$ gauge group. One can not treat it as an $SO(3)$ gauge group because the doublets are not bifundamentals of $SO(3)$.

Because of the need for bifundamental matter, certain exotic representations are not allowed. Many of the quantum numbers for possible resonant particles, as in  \cite{Han:2010rf}, are forbidden.

\subsubsection{Light $Z'$'s:}

It is possible to argue that a single $Z'$ is the minimal content beyond the Standard model that  is stringy in this sense \cite{Berenstein:2006pk}. This is called the Minimal Quiver Standard Model (MQSM) and this can be analyzed in the absence of supersymmetry. Notice that when strings are a TeV scale, one can have supersymmetry breaking at the string scale because one does not need a second solution to the hierarchy problem to be in place. 
The corresponding gauge theory is a three stack $U(3)\times Sp(1) \times U(1)$ model. One linear combination of the $U(1)$ is anomaly free and gives the hypercharge. The other linear combination is anomalous and becomes massive via a St\"uckleberg mass which is required by the Green-Schwarz anomaly cancellation mechanism. The model has no dark matter candidate. This would probably need to be supplied by Kaluza-Klein excitations of the closed strings, so it would not be a Weakly Interacting Massive Particle.

As far as effective field theory goes,  the couplings of this $Z'$ can be completely computed , but the mass of $Z'$ is not determined. This is of order $g_o M_s$. In this case, the $Z'$ does not generate Flavor Changing Neutral Currents (FCNC's) and it does not mix much with the $Z$, so minimal precision constraints are met and one can bound
the mass of the $Z'$ in this model with collider data, as well as using bounds on the $\rho$ parameter \cite{Berenstein:2008xg}.  In this model it turns out  that the $Z'$ is leptophobic and couples
to quarks with a coupling constant related to the strong interactions. Three stack models are analyzed in \cite{Cvetic:2011iq}.

A four stack model is analyzed in \cite{Anchordoqui:2011eg}, where one of the $Z'$ arises from a $B-L$ anomaly free setup. These extensions can be done in supersymmetric setups  \cite{Anastasopoulos:2008jt} , but notice that D-brane setups only allow for specific values of the charges of the extended $U(1)$ symmetry, as well as diagonal kinetic terms, restricting various mixings. The collider phenomenology of a single such an anomalous $Z'$ boson (the lightest one that would be accessible in a more general model) can be studied in detail \cite{Kumar:2007zza}. A rather comprehensive review of the physics of $Z'$ gauge bosons can be found in \cite{Langacker:2008yv}. 

Other models tend to have more $U(1)$ factors, leading to additional $Z'$ bosons. These can mix with each other, so bounds from direct detection are weaker, since they depend on more parameters. On the other hand, these can induce FCNC's depending on how one arranges the model. In such cases, the indirect bounds from precision measurements are more stringent. Models such as those have been analyzed in \cite{Coriano':2005js} and lead also to interesting corrections to various vertices of gauge bosons. These are tied to the anomaly.

\subsubsection{Multi-Higgs fields: }

An interesting possibility arises in the case that one enhances the $SU(2)$ weak to a full $U(2)$, because one might then be able to discriminate one generation of quarks and distinguish it from the other two: the $( 3,\bar 2)$ of $U(3)\times U(2)$ is different than the $( 3,  2)$ \cite{Ibanez:2001nd}. If at each $U(N)$ brane stack there are only fields in the fundamental or antifundamental (no 2-tensors), then cancellation of tadpoles requires the same number of fundamentals as antifundamentals. This is true even for a $U(1)$ or $U(2)$ brane. 
Tadpole cancellation at the $U(2)_w$ brane would have the quarks in two copies of $(3,2)$ and one copy of $(3, \bar 2)$, and the leptons in three copies of the $(1,\bar 2)$ as far as the $U(3)\times U(2)$ quantum numbers are concerned. The $U(1)$ in the $U(2)_w$ in this case can not participate in hypercharge: it would give a different hypercharge assignment to one quark doublet versus the other two. Also, a Higgs field which is charged under $SU(2)$ would also have a charge under the $U(1)\subset U(2)$, apart from hypercharge. In such a setup one would require at least two different Higgs fields. One of them would give masses to two up quarks
and one down quark, and the other would give masses to two down quarks and one up quark. If the model is supersymmetric, then the spectrum would get doubled again (if one is to generate all quark masses from superpotential Yukawa couplings). Such models distinguish between generations and generally lead to disastrous FCNC violations from $Z'$ exchanges unless the string scale is high $M_S> 10^5 TeV$, furthermore, scalar box diagrams from the multi-Higgs fields would also lead to FCNC operators. This pushes the scale of these fields high.

\subsubsection{Axions: }In  a non-supersymmetric setup even though one would need only two different types of Higgs fields to get all the Standard Model couplings from perturbation theory, they are still sensitive to flavor. They are not a general two Higgs doublet model: the perturbative potential is constrained by the existence of a global $U(1)$ symmetry which is broken when the Higgs acquires a vacuum expectation value. This is the left-over $U(1)$ global symmetry after the $Z'$ becomes massive, as described in the Iba\~nez-Quevedo mechanism.
This would generically lead to a goldstone boson whose mass would need to be generated non-perturbatively, and this would not be an invisible QCD axion: such a goldstone boson would be associated to the electroweak theory. Another possibility is to add more particles: an antisymmetric scalar of $U(2)$ might do such a trick.  It is not charged under the $SU(2)_w$. Such a scalar vev would break $U(2)\to Sp(1)$ and would make the $U(1)$ massive. Because the $U(1)$ gauges some family symmetry, it can be considered as a theory of flavor. Such theories that arise from gauging family symmetries typically lead to axions that arise from the open string sector \cite{Berenstein:2010ta,Berenstein:2012eg}. Such axions, if they have a low axion decay constant, are in conflict with observational bounds (for a review see \cite{Turner:1989vc}). The axion decay constant is  lower than the string scale in these models, so for the models to be consistent, the string scale gets pushed up above $10^9$ GeV. One can also have trouble in such models with the hierarchy of quark masses, because they are controlled by a Froggat-Nielsen mechanism, and the pattern in the models does not necessarily reflect the observed pattern in nature. Simple models as these have trouble describing the CKM matrix.  Also the axion photon-photon coupling is different than the ones that arise from Grand Unified Theories, which can make them more easily detectable. Supersymmetric setups which lead to such axions have also been analyzed in \cite{Coriano:2006rg}. 

To have a realistic  axion survive in a low string scale scenario, it needs to arise from the closed string state that is participating in the anomaly, rather than the open string states.
Such a particle, if it is very light, is still sensitive to axion bounds.
In multi-Higgs models, even in the absence of such an axion,  one also generically ends up in trouble with FCNC bounds and thus one would like to push all masses (except the Standard Model Higgs) to higher scales, way beyond the TeV scale. The reader interested in axion physics is encouraged to read the paper on axions in string theory \cite{Svrcek:2006yi}.

\subsubsection{Neutrinos:} Another problem with models at the TeV scale is that neutrino masses tend to be too large. A see-saw mechanism would not render the appropriate smallness of the neutrino masses. This problem can be solved by insisting that Lepton number is conserved (in perturbation theory)  and not having a right handed neutrino readily available  on the brane \cite{Antoniadis:2002qm}, if the right handed neutrino lives in the bulk, then it's coupling to the neutrino in the brane is suppressed by volume effects. Other possibilities can be found in \cite{Cvetic:2008hi,Cvetic:2010mm}.

\subsubsection{Tadpole cancellation and Exotics}

There are some predictions of top-down models that can not be attributed to low energy physics like cancellation of anomalies. This is the case where tadpole cancellation requires the introduction of an additional D-brane
in the spectrum that does not connect directly to the Standard model. This is, the messengers that connect the Standard model to the new sector are non-chiral with respect to the Standard model. Such model can have a sequestered low energy spectrum that might be a good candidate for dark matter. A recent example has been analyzed in \cite{Cvetic:2012kj}. The main observation is that in an $U(2)$ node there is no $SU(2)$ anomaly, but the $U(2)$ stack might not cancel tadpoles. Therefore one would need to introduce additional matter, made of open strings,  charged under the $U(1)$ which might end on another brane that would furnish the dark sector. These setups can give rise to fractionally charged matter that is not colored. One can also get a hidden valley scenario for dark matter.

In general, top down models can predict a lot of additional particles beyond the Standard model. If they are chiral and charged under the electroweak theory, they get their mass from electroweak symmetry breaking. Such particles are essentially ruled out by accelerator searches and such models should be considered to be ruled out. On the other hand, when such matter is vector-like, it is allowed. This is well documented in other reviews (see for example \cite{Blumenhagen:2005mu}).  When strings are at a TeV scale, it is hard to distinguish such states from other string resonances, so we will not pursue this issue further.

\subsection{String scale bounds}

So far, we have argued that the string scale is not an output of specific constructions and that it generically acts as a free parameter. Certain field theory arguments (what one can call indirect detection signals, or absence of them, like in FCNC's) can push the string scale to be high, but it is also useful to understand what the direct limits on string state searches are and how they differ from other searches. This is usually handled at the level of a toy model: we make no attempt at seriously understanding issues related to flavor, axions, dark matter, and instead focus on the idea that string theory predicts stringy resonances with various spins and forming Regge trajectories. The prototypical setup is the one pioneered in \cite{Cullen:2000ef} (see also \cite{Burikham:2004su}): we dress a familiar process like $e^+e^-\to \gamma\gamma$ with some stringy information related to the presence of these additional resonances.This usually results in mutiplying a familiar amplitude by the  Veneziano amplitude.
This is
\begin{equation}
A(1,2,3,4)_{stringy} \simeq  {\cal S}(s,t) A(1,2,3,4)_{SM}
\end{equation}
and 
\begin{equation}
{\cal S}(s,t) \simeq \frac{\Gamma(1-\alpha's) \Gamma(1-\alpha' t)}{\Gamma(1-\alpha't-\alpha' s )}
\end{equation}
The poles in the $\Gamma$ function we're interested in arise in the $s$-channel and give a peak in the cross section of the particle physics experiment. If we're not able to get to the string scale directly, then we do a Taylor series in $s,t$ around $s,t\simeq 0$ and we get effective field theory corrections to an amplitude that is well known. These corrections are polynomial in $s,t$, so they grow with the energy and would modify the tails of distributions for high energy collisions in a proton collider. 

A more recent analysis \cite{Anchordoqui:2007da} considers other two to two scattering processes with the LHC collider in mind. In particular, the initial state is taken to be a pair of quarks, a pair of gluons, or a quark and a gluon. In particular, in that paper, they consider looking into the process $gg\to g \gamma$ where two gluons fuse and in the final state there is one gluon and a photon. Since as described previously, the $SU(3)$ gets enhanced to $U(3)$, a single trace operator with four insertions can have the structure $\hbox{tr} ( G_{\mu\nu}^4)\simeq \hbox{tr}((G_{SU(3)})^3_{\mu\nu})G_{\mu\nu}{U(1)}$ where there are three external gluons and one extra neutral vector boson particle. This one generally participates in hyperchagre, so it can be either a $Z$ or a photon. This is a $pp\to \gamma+jet$ topology, which can also be used to detect some mixing angles of neutral gauge bosons \cite{Anchordoqui:2008ac}.

The other standard topologies to look for these resonances are $gg\to gg$ and $gq\to gq$, so a resonance can be looked for in the dijet invariant mass spectrum.
These ideas are reviewed in \cite{Lust:2008qc,Anchordoqui:2009mm}.

It is important to notice that in the $qq\to qq$ channel, states that are not just the Regge trajectories of the gluon can appear. If there is a spectrum of Kaluza-Klein excitations of the gluon, it could become available in this channel. If there is a notion of Kaluza Klein (KK) parity,  such KK odd parity states are produced in pairs and the lightest KK odd state could be a candidate for dark matter. Such reactions would lead to missing energy and are captured by other searches.

Direct bounds from these reactions are roughly $M_s>4-5 TeV$ (see for example \cite{Chatrchyan:2011ns,ATLAS:2012pu}) and is quoted as $M_S>4.78$ TeV in the recent work \cite{Lust:2013koa}. Notice that these are based on a toy model of string theory at the LHC scale, where one just mutiplies a known amplitude by a standard kinematical  factor that appears in many models with flat branes. In a realistic model one can expect that details will matter much more. In particular, one would expect that the different spin components would acquire different masses. 
A direct search bound for $Z'$ tends to push the string scale higher, but the bound is lower, $M_{Z'} > 1.62TeV$ (assuming the same couplings as a $Z$). This pushes the string scale up because the mass of the (anomalous) $Z'$ is suppressed relative to the string scale by a factor of the open string coupling constant. General models with resonances in these channels can be analyzed as in \cite{Han:2010rf} to arrive at different (more general) bounds.

To go further along this line of thought requires also the understanding of $2\to 3$ processes. The technology to deal with these amplitudes is developed in \cite{Lust:2009pz}.
One can also study $\nu q\to \nu q $ scattering and $\nu g \to \nu g $ scattering (see \cite{Friess:2002cc} and references therein)  and obtain bounds from cosmic neutrinos, finding a reach of about $1 TeV$.
One should point out that in quiver constructions, the neutrino is located on a different stack of branes than the gluon, so the effective terms in the action that can give rise to a $\nu g \to \nu g $ reaction are double trace operators, and the amplitude should be suppressed by an extra factor of $g_c$, limiting further the reach of such searches.

\section{CLOSING REMARKS}

At this point, it is perhaps appropriate to editorialize some sentiments (not exactly predictions) of the author with regards to the prospects of having string physics with a  string scale at the TeV scale. 
It is important to note that while strings at the TeV scale are not ruled out by experiments, the case for them is not particularly good. In particular, one expects that a theory with a lot of states at the TeV scale will run into trouble with flavor constraints: these particles will participate in loops and produce violations of flavor physics. Indeed, they would be resonances of flavored quarks and they would carry flavor information.
A serious attempt to do a theory of flavor with strings at the TeV scale is compulsory if one is to explain the absence of flavor-changing neutral currents, and what is the mechanism responsible for their cancellation. Toy models of theories of flavor (as in \cite{Berenstein:2010ta}) seem to run into trouble with axions, so they push the string scale to be much higher, of order at least $10^{10}$ GeV where the flavor bounds don't matter anymore: all particles are too heavy. Furthermore, one would expect that such a theory could also account for inflationary cosmological data and the matter anti-matter asymmetry. 
The author has not seen any convincing cosmological model where the scale of new physics can be that low. One should furthermore be warned that a string spectrum leads to a Hagedorn transition right around the string scale and it is not clear what would be the cosmological signature of such a phase transition, or how that phase transition could limit cosmological models. In regards to axions, it should be very interesting to have at least one such axion to solve the QCD strong CP problem, but it's decay constant would be much larger than the string scale. Not all axions have this property and understanding how to avoid this problem (without sending all scales higher) has not been addressed sufficiently in the literature. 
Finally, any such model should have a serious candidate for dark matter and its associated phenomenology and in particular, it should tell us if it is a WIMP, a hidden valley, or perhaps even an axion.

\section*{Acknowledgements}
 The author would like to thank J. Halverson and P. Langacker for a critical reading an earlier version of this manuscript. Work supported in part by DOE under grant DE-FG02-91ER40618.

\end{document}